\documentclass[%
reprint,
 amsmath,amssymb,
 aps,
]{revtex4-1}

\usepackage{graphicx}
\usepackage{color}
\usepackage{dcolumn}
\usepackage{bm}

\begin{document}


\title{Reproducible Operating Margins on a 72,800-Device Digital Superconducting Chip}
\author{Quentin P. Herr, Joshua Osborne, Micah J.A. Stoutimore, Harold Hearne, Ryan Selig, Jacob Vogel, Eileen Min, Vladimir V. Talanov, and Anna Y. Herr}
\affiliation{Northrop Grumman Systems Corp., Baltimore, Maryland 21240, USA}
\email[email: ]{quentin.herr@ngc.com}
\thanks{\\This research is based upon work supported in part by the Office of the Director of National Intelligence (ODNI), Intelligence Advanced Research Projects Activity (IARPA), via the U.S. Army Research Office, contract number W911NF-14-C-0116. The views and conclusions contained herein are those of the authors and should not be interpreted as necessarily representing the official policies or endorsements, either expressed or implied, of the ODNI, IARPA, or the U.S. Government. The U.S. Government is authorized to reproduce and distribute reprints for Governmental purposes notwithstanding any copyright annotation thereon.
}

\date{2 June 2015}

\begin{abstract}

  We report the design and test of Reciprocal Quantum Logic
  shift-register yield vehicles consisting of up to 72,800 Josephson
  junction devices per die, the largest digital superconducting
  circuits ever reported. Multiple physical layout styles were matched
  to the MIT Lincoln Laboratory foundry, which supports processes with
  both four and eight metal layers and minimum feature size of
  0.5\,$\mu$m. The largest individual circuits with 40,400 junctions
  indicate large operating margins of $\pm$20\% on AC clock
  amplitude. In one case the data were reproducible to the accuracy of
  the measurement, $\pm$1\% across five thermal cycles using only the
  rudimentary precautions of passive mu-metal magnetic shielding and a
  controlled cool-down rate of 3\,mK/s in the test fixture. We
  conclude that with proper mitigation techniques, flux-trapping is no
  longer a limiting consideration for very-large-scale-integration of
  superconductor digital logic.

\end{abstract}


\maketitle


Superconductor digital technology offers fundamental advantages over
conventional semiconductor technology in terms of power efficiency,
interconnect bandwidth, and computational density, but to realize this
potential the integration scale must increase. Past limitations to
scaling have included 1) design, as dc-powered circuits based on Rapid
Single Flux Quantum (RSFQ) logic draw 1\,A per 1,000 gates, 2)
fabrication, as non-planarized processes allow only four metal layers
and feature sizes greater than 1\,$\mu$m
\cite{abelson2004superconductor}, and 3) test, as flux trapping in the
superconductor films can degrade or preclude correct circuit
operation. Scaling superconductor technology is now possible due to
recent advances in circuit design embodied in Reciprocal Quantum Logic
(RQL) \cite{herr2011ultra, herr20138} and recent advances in
superconductor integrated circuit fabrication, which extends to
minimum features of 0.25-0.5\,$\mu$m and 6-8 levels of metal at
multiple foundries \cite{johnson2010scalable,
  tolpygo2015fabrication}. This paper addresses flux trapping as the
one remaining technological obstacle limiting integration scale. We
measure flux-trapping signatures in large RQL shift register circuits
and report physical layout styles and test protocols that completely
eliminate the effect.

\section{Flux Trapping in Superconductor Integrated Circuits}

Flux trapping quantizes and localizes magnetic field as
single-flux-quantum (SFQ) current vortices in the superconductor films
as they are cooled through the transition temperature. Earth's ambient
field of about 40\,$\mu$T would generate a magnetic flux of 1\,nTm$^2$
through the surface of a 5\,mm-square chip. Since the SFQ is
$\Phi_0=h/2e \approx 2.07 \times 10^{-15}$\,Wb$=2.07 \times
10^{-15}$\,Tm$^2$, this amounts to about 500k trapped vortices. Vortex
radius is defined by the London penetration depth $\lambda_L \approx
0.1$\,$\mu$m in Nb at 4.2\,K. The magnetic field local to the vortex
corresponds to the critical field for Nb and is larger than the earth
ambient by a factor of 10$^3$. Note that a reduction in ambient field
will result in a proportionate decrease in the number of trapped flux,
but will not change the magnitude of an individual trapped flux.

Flux trapping has been observed directly as a shift in the threshold
characteristic for simple two-junction Superconducting-Quantum-
Interference-Device (SQUID) circuits \cite{bermon1983moat,
  nagasawa1995evaluation, fujiwara2009research} and even more directly
using magnetic imaging \cite{jeffery1995magnetic}. For larger, digital
circuits, flux trapping produces reduced operating margins or
non-functional circuits \cite{robertazzi1997flux}. Flux trapping is
stochastic. The hallmark of flux trapping is that all of the above
observables vary from one-to-the-next thermal cycle through the
superconducting transition.

Standard mitigation of flux trapping in the test fixture involves two
or three concentric mu-metal shields to reduce earth ambient field by
about a factor of 40 down to 1\,$\mu$T. Additional precautions include
1) avoidance of thermal gradients using slow cool-down rates, achieved
using a thermometer and heater (or closed-cycle refrigerator) running
in a control loop, 2) reduction of residual field using in-situ
demagnetization of the mu-metal shields while cold, and 3) active
field cancellation of residual magnetic fields using feedback. See the
above references \cite{bermon1983moat, nagasawa1995evaluation,
  fujiwara2009research, jeffery1995magnetic, robertazzi1997flux} for
examples of each of these precautions. Applying all of these
techniques at once can reduce the residual field another two orders to
10\,nT \cite{polyakov20113d}, which amounts to only 100 vortices
through the chip.

Mitigation of flux trapping in the physical design involves holes in
the ground plane, which provide energetically-favorable sites to
sequester trapped flux. Holes with high aspect ratio called moats give
the best results \cite{bermon1983moat, jeffery1995magnetic,
  robertazzi1997flux}, but a perforated-moat geometry is nearly as
good as a continuous moat \cite{nagasawa1995evaluation,
  fujiwara2009research}. While helpful, these precautions have not
proven to be fully effective, thereby limiting the integration scale
to an estimated 10,000 Josephson junctions
\cite{terai2007diagnostic}. A more pessimistic result was reported in
\cite{semenov1999extraction}. However, this group has made continuous
progress in both design and test \cite{polyakov2007flux,
  narayana2009evaluation, polyakov20113d} and has reported a quite
favorable result at the 0.5\,$\mu$m node for a circuit with 32,800
junctions having only a few outliers attributed to flux trapping
\cite{semenov2015new}.

Reported mitigation of flux trapping in integrated circuits has been
inconclusive at best. However, simple well-controlled experiments
indicate that patterning a single layer into strips can be fully
effective for both low-temperature and high-temperature
superconductors \cite{stan2004critical, kuit2008vortex}. No flux
trapping was observed in 200-nm-thick, 15\,$\mu$m-wide Nb strips in
ambient field up to 10\,$\mu$T with a cool-down rate of 10\,mK/s
\cite{stan2004critical}. At temperatures below the critical
temperature, $T_c$, but above the vortex freezing temperature $T_f
\approx T_c-15$\,mK, due to $\lambda_L \gg d$ the vortex radius is
defined by the Pearl length $\Lambda = 2\lambda_L^2/d$, with $d$ the
film thickness \cite{pearl1964current}. Thus, for moat spacing
$W=2\Lambda(T_f)\approx 20$\,$\mu$m the vortex is sure to be attracted
by the image anti-vortex towards the moat edge (and eventually
sequestered there) if the ambient field is kept under
$\Phi_0/W^2$. This is the critical field for complete vortex expulsion
from a narrow strip of width $W \ll \Lambda$ \cite{clem1998vortex,
  likharev1972formation, maksimova1998mixed, kuit2008vortex}. We apply
this length scale to moat geometries in integrated circuits at
sub-micron.

\section{Yield Vehicle Design}

\begin{figure}
\includegraphics[width=3.4in]{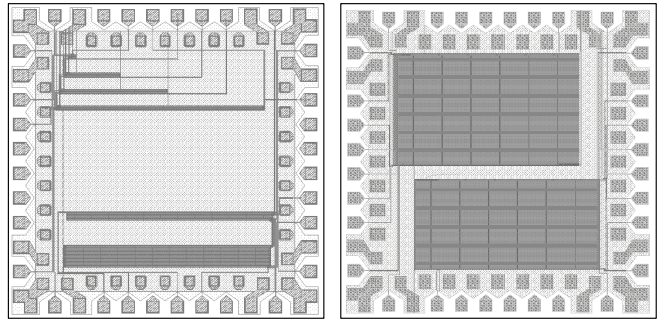}
\caption{ {\bf CAD drawings of two 5\,mm-square chips}
show designs with the lowest and highest junction
count. Each chip contains two circuit blocks powered with independent
clock lines. Each block has one or more shift registers sharing a
common input and having separate outputs. The chip with the larger circuits
contains two independent shift registers of 32,400 and 40,400
Josephson junctions that fill the 3\,mm-square active area.
\label{fig1}}
\end{figure}

We designed and tested yield vehicles consisting of RQL shift
registers with eight Josephson junctions powered by a four-phase AC
clock that are triggered sequentially by RQL-encoded data to produce
one clock cycle of delay. An exponential progression started with
small circuits of just a few stages and moved up to long serpentines
that filled the chip (Fig.~\ref{fig1}). Such a simple design does not
allow faults to be isolated within the circuit, but is adequate for
measuring the characteristic maximum size of functional circuits.  The
chips were designed with density approaching 100,000 devices on a $5
\times 5$\,mm die for the SFQ3ee and SFQ4ee integrated circuit
processes at the MIT Lincoln laboratory, which represent
state-of-the-art superconducting fabrication
\cite{tolpygo2015fabrication}. However, these processes are only the
initial steps on a road-map to much higher densities at more advanced
lithography nodes \cite{tolpygo2014roadmap}.

\begin{figure}
\includegraphics[width=3in]{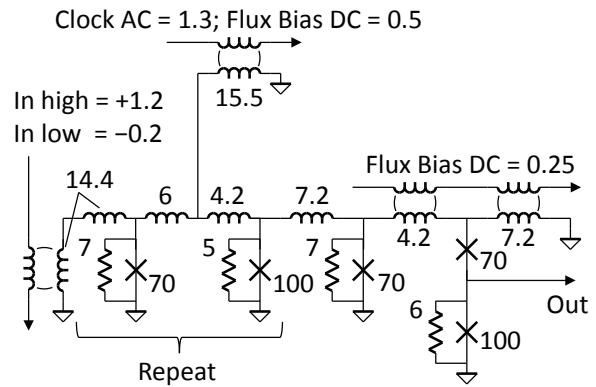}
\caption{ {\bf The yield vehicle schematic} consists of an
edge-triggered input that converts a pattern to RQL
data encoding, multiple ac-powered shift register stages,
and an output that converts junction phase to an observable voltage.
Four repeated stages powered with a four-phase clock produce one clock
cycle of delay. The total number of repeated stages in each circuit ranged
from less than 10 to greater than 20,000. Parameter values are shown
with units of $\mu$A for the junction critical currents, pH for the
inductors, $\Omega$ for the resistors. Input signals to the transformers
are given in units of $\Phi_0$, equal to the product of current in the
primary and mutual inductance. The output is dc-biased at 130\,$\mu$A and
produces a peak-to-peak voltage of 0.5\,mV for a target junction critical
current density of 100\,$\mu$A/$\mu$m$^2$. 
\label{fig2}}
\end{figure}

The circuit schematic (Fig.~\ref{fig2}) is similar to that reported in
\cite{herr2011ultra}. Most of the circuits use junctions with critical
currents of 70-100\,$\mu$A, which is half of that previously
reported. Only the SFQ3ee four-metal-layer design used the original
junction critical currents of 140-200\,$\mu$A. The output circuit is
conceptually similar to the SFQ-to-DC converter
\cite{likharev1991rsfq} but is compatible with RQL data encoding. The
output has only three Josephson junctions and produces 0.5\,mV, which
is adequate for the intended sampling measurement.

Three physical layout styles were developed for two different versions
of the fabrication process, SFQ3ee and SFQ4ee, at the MIT Lincoln
Laboratory \cite{tolpygo2015fabrication}. The two processes have
similar feature size but a different number of metal layers. The
SFQ4ee process has eight metal layers, M0-M7, while the SFQ3ee process
has only four metal layers, M4-M7, corresponding to the topmost layers
in the SFQ4ee stack-up. Our various layout styles differ primarily in
the choice of ground plane layers in the physical layout.

\begin{figure}
\includegraphics[width=3.0in]{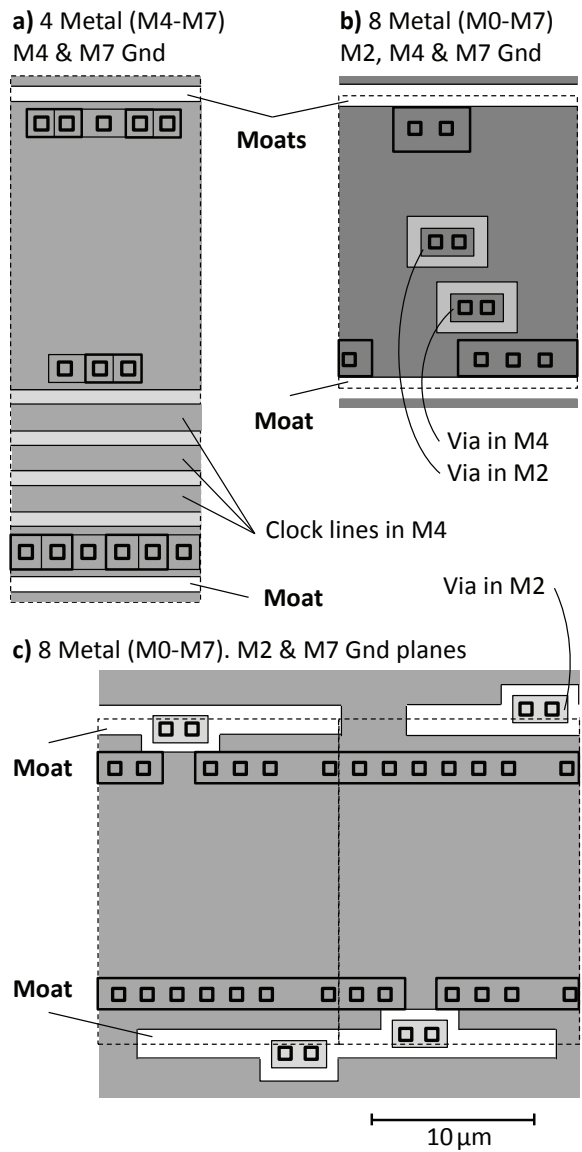}
\caption{ {\bf Moat geometries and ground planes} are shown for
  three physical layout styles, which we will refer to as a)
  one-and-a-half ground planes, b) three ground planes, and c) two
  ground planes. Light-shaded regions indicate a single layer of
  ground metal and medium shade indicates two layers. The darkest
  shade, used in (b) indicates three layers of ground metal. Coincident moats
  in all ground layers are unshaded. Dashed
  lines indicate the unit cells. Unit cell size for the three layouts
  is 11.7\,$\mu$m$\times$32.7\,$\mu$m,
  14.8\,$\mu$m$\times$18\,$\mu$m, and
  14.8\,$\mu$m$\times$20\,$\mu$m. Two unit cells are shown in (c) in
  order to capture the moat geometry, which spans two cells.
  \label{fig3}}
\end{figure}

The first layout style uses the SFQ3ee process and is similar to that
reported in \cite{herr2011ultra}, but with feature size scaled down to
sub-micron design rules. The AC clock lines and bias
transformers were laid out beside the Josephson junctions and the
interconnect inductors. The active region used two ground planes, M4 and
M7, which are the top and bottom metal layers in the
stack-up. However, a single ground plane on M7 was used over the clock
lines, which were patterned in M4. As this layout style has a mix of
single and double ground planes, we will refer to it as having
one-and-a-half ground planes. Where two ground planes are present,
coincident moats were patterned in both layers. Ground metal for this
style is shown (Fig.~\ref{fig3}a).

The second layout style, using the SFQ4ee process, had three global
ground planes with coincident moats laid out in M2, M4, and M7
(Fig.~\ref{fig3}b). More metal layers allowed increased vertical
integration. The AC clock lines and bias transformers were laid out in
the M0 and M1 metal layers, under the Josephson junctions and the
interconnect inductors that used M3, M5, and M6. Additional features
were patterned in the M2 and M4 ground planes to accommodate
thru-vias, which were intentionally staggered.

The third layout style (Fig.~\ref{fig3}c) also used the SFQ4ee process
but had only two ground planes, M2 and M7. Thru-vias were placed in
the moats. To get above the M2 ground plane, a wire in M1 extended to
the via in the moat and a wire in M3 followed the same path back over
the ground plane.  Coupling from a flux trapped in the moat into the
M1/M3 loop would be small as the loop is orthogonal to the
moat. Extensive via walls around the moat shield the current
associated with trapped flux from the active circuit. Another
significant change from the previous layout style is that the moats in
this design are only 26\,$\mu$m long and are separated by a 3.6\,$\mu$m
gap, instead of being continuous structures with a length scale
similar to the dimensions of the circuit, about 3\,mm. The shorter
moats with gaps are more amenable to the X-Y interconnect needed for
more complex logic circuits.

\section{Test}

Chips were mounted in a pressure-contact probe with three concentric
mu-metal shields to attenuate ambient field below 1\,$\mu$T, which
is an order of magnitude less than the critical field for complete
vortex expulsion, $\Phi_0/W^2\approx 10$\,$\mu$T for our typical moat
separation, $W\approx15$\,$\mu$m. The probe was lowered into an LHe
transport dewar to achieve the 4.2\,K operating temperature. All chips
were tested using a manual measurement in which the cooling rate
through the transition temperature was neither observable nor
well-controlled, but is estimated to have been 0.1-10\,K/s.  Circuits
were tested using a simple repetitive bit sequence from a pattern
generator connected to chip input, and output was observed on a
sampling oscilloscope after passing through a low-noise amplifier.
For convenience, the tests were conducted at a 2\,GHz rate, which is
much lower than the intrinsic maximum speed of the shift
registers. Operating margins on clock power where measured for
functional circuits by visually matching the output to the expected
bit sequence. The point of circuit failure was somewhat subjective,
but is estimated to be accurate to $\pm$0.2\,dB as the onset of
errors is quite rapid.

The largest circuits were retested using an automated measurement with
improved test procedures: 1) The analog waveforms from the sampling
oscilloscope were downloaded to a PC and digitized using a simple
threshold algorithm. Operating margins on clock power where measured
using an automated binary search that both compared the digitized
output to the expected pattern, and set the clock power produced by
the sine-wave generator. 2) The cool-down rate through the transition
temperature was controlled using a thermometer and a heater to be
about 3\,mK/s. This proved to be very important.

\begin{figure}
\includegraphics[width=3.4in]{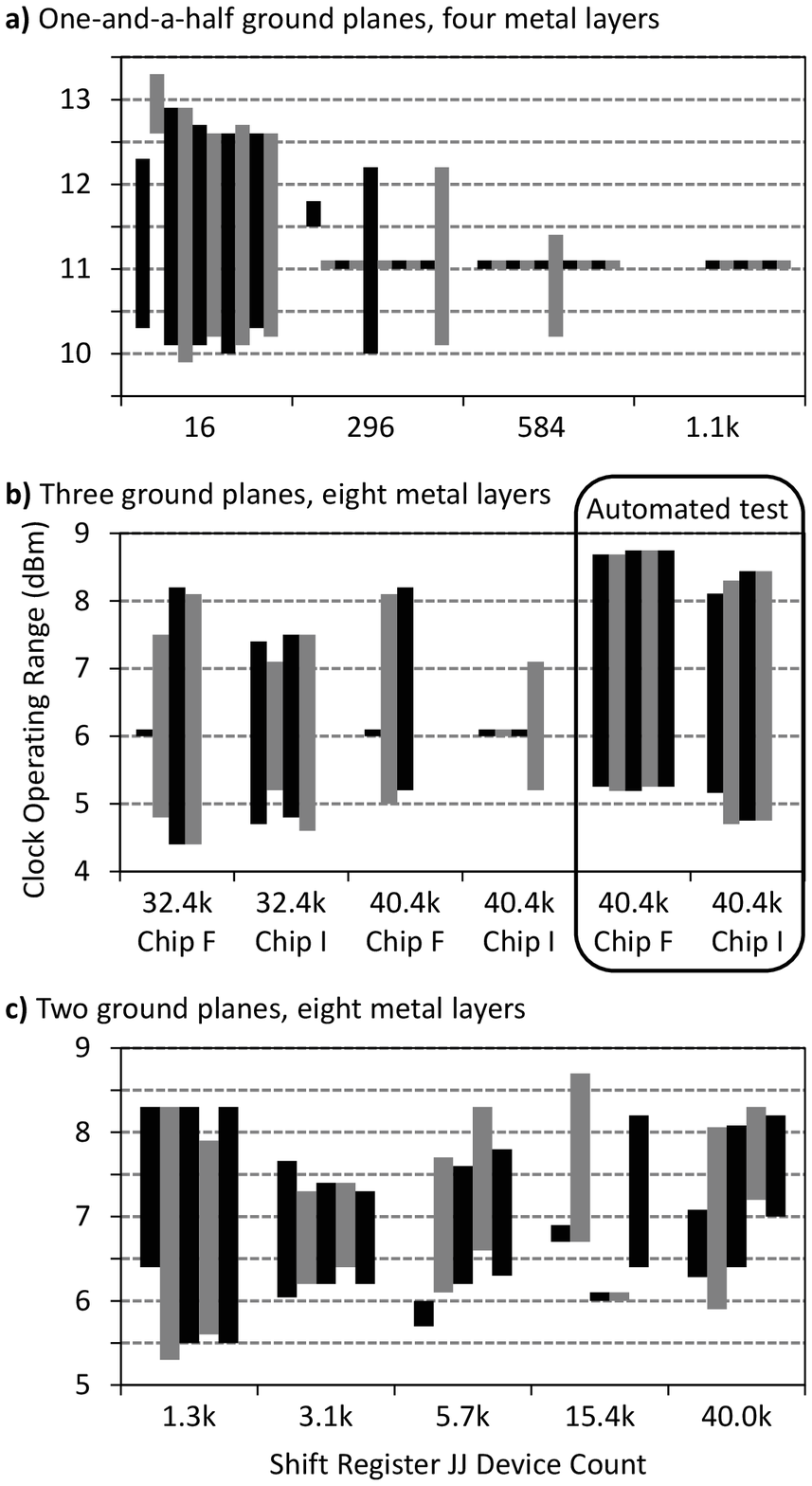}
\caption{ {\bf Shift-register yield vehicle circuits} using the three
  different layout styles were characterized across multiple thermal
  cycles. The bar plots indicate the measured operating margins on the
  clock power. The smallest bars, plotted as 0.1\,dB, are placeholders
  indicating an attempted measurement with no operating point
  found. The power scale is arbitrary, corresponding to the source
  before various levels of attenuation. In all cases, the design value
  for power on-chip was $-$14.5\,dBm, corresponding to an amplitude of
  1.2\,mA on a 50\,$\Omega$ line.  a) Typical operating margins for a
  chip with one-and-a-half ground planes. Multiple chips across several
  fabrication runs were measured with similar results. b) Operating
  margins for two chips with three-ground-planes, representative of
  the best and worst measurements of eight chips from one
  wafer. Retest of the larger circuits using an automated measurement
  with a slow cool-down rate shows little or no evidence of flux
  trapping. c) Operating margins for a single chip with
  two-ground-planes, with the largest circuit comparable in size to
  the largest circuit in (b).
  \label{fig4}}
\end{figure}

For the one-and-a-half-ground-plane layout style, representative
results from one chip are shown in Fig.~\ref{fig4}a. On subsequent
cooldowns, the 16-junction circuit worked with margins of 2-3\,dB
with 90\% probability, which is sufficient to validate the circuit
schematic, fabrication process, and test fixture. However, the
probability of a working circuit fell off rapidly with increasing
circuit size. The 296-junction circuit showed similar margins with
only 20\% probability, the 584 junction circuit worked only 10\% of
the time, with reduced margins, and the 1,136-junction circuit was
found to be nonfunctional in six cool-down attempts. This result
indicates high flux-trapping incidence compared to a
previously-reported circuit of similar junction count and similar
layout style, but fabricated with 2\,$\mu$m minimum feature
size. Without more data this result would indicate that flux-trapping
incidence increases as feature size decreases to sub-micron.

For the three-ground-plane layout style, multiple chips from two
wafers were tested. In stark contrast to the previous result, shift
registers of all sizes were found to be functional. Test effort was
directed to the chips containing the two largest circuits, with 32,400
and 40,400 Josephson junctions, which together filled the
3\,mm$\times$3\,mm active area of the chip. Fig.~\ref{fig4}b shows
results for two chips, labeled F and I, that are representative of the
most favorable and least favorable data collected. For the manual
test, operating margins are not reproducible from one cool-down to the
next, but only a few attempts are needed in order to establish wide
margins. The widest margins are only weakly correlated to circuit
size. Taken together, the two circuits on Chip F represent a
72,800 Josephson junction chip with operating margins of 3\,dB. This is the
largest digital superconducting chip ever reported.

Re-test of the larger circuits using the automated setup is also shown
in Fig.~\ref{fig4}b. For Chip I, the circuit was found to be
functional on all four cool-downs with relatively reproducible margins
ranging from 3-3.7\,dB. For Chip F the circuit was functional on all
five cool-downs with margins that were reproducible to within
$\pm$0.06\,dB, the resolution of the automated binary search. This
corresponds to reproducibility of clock amplitude of about
$\pm$1\%. As other factors such as system noise or variable contact
resistance in the pressure contact probe could account for this level
of variability between cool-downs, the result produced no evidence of
flux trapping.

Finally, for the two-ground-plane layout style, data for a single chip
collected using manual test are shown in Fig.~\ref{fig4}c. The largest
circuit, comparable in size to the largest three-ground-plane
circuits, was functional on five-of-five cool-downs but with varying
operating margins. Overall the data indicate that this layout style
had roughly similar effectiveness at sequestering trapped flux in the
moats as the three-ground-plane style.

\section{Discussion}

Our main result is an existence proof of large RQL circuits of up to
72,800 Josephson junctions per die fabricated in an eight-metal-layer
sub-micron process operating with large operating margins and showing
near-immunity to flux trapping. A circuit with 40,400 Josephson
junctions was characterized across five thermal cycles with no
evidence of flux trapping, using only the rudimentary precautions of
passive mu-metal magnetic shielding and a controlled cool-down rate of
3\,mK/s in the test fixture. The relevant metric for flux-trapping
statistics in integrated circuits is not junction count per se but
active area. We report immunity to flux trapping for circuits with an
active area of up to 3\,mm square. We expect this result to hold for
future circuits with higher density and higher junction count
occupying a similar physical area.

We also report very poor performance of circuits fabricated in the
four-metal-layer version of the process using one-and-a-half ground
planes, which we attribute to flux trapping. Only circuits of less
than 600 junctions were found to be functional. By comparison, we
previously reported operation of a 1,200-junction circuit with a
similar layout style fabricated in a non-planarized, 2\,$\mu$m process
\cite{herr2011ultra}. We developed a simple narrative to account for
these poor initial results, under the assumptions that the ground
planes may have slightly different transition temperatures, and that
the ground planes are effectively superconducting before finer
features such as vias and wires. If the M7 global ground plane goes
superconducting first and sequesters a trapped flux in the moat, it is
plausible that the trapped flux will also find the corresponding moat
in the M4 half-ground-plane. However, if the M4 half-ground-plane goes
superconducting first, the trapped flux may instead be expelled to the
slits that are used to define the clock lines. Subsequent trapping in
the moat in M7 will cause magnetic field to thread through the active
region on the circuit. In this scenario, the moats concentrate field
exactly where it is not wanted, and may be worse than no moats at all.

The technological solutions that produced flux-trapping immunity in
the largest circuits fall into three general categorizes: 1) the
eight-metal-layer, planarized, sub-micron fabrication process, 2) the
moat geometry in physical layout, and 3) the slow, controlled
cool-down in circuit test. These solutions are expected to be general
to all superconductor integrated circuits and do not depend on the
particulars of the RQL circuits reported here. The eight-metal-layer
process affords vertical integration that eliminates the compromises
in layout style that were present in the four-metal-layer
design. Planarization avoids film topology that could produce
undesirable pinning centers for vortices in the ground plane. Physical
layout was centered around moat design done in accordance with the
analysis presented in Section I. Long moats with spacing not greater
than 15\,$\mu$m were designed to produce complete vortex expulsion
from the patterned ground plane. Continuous moats in the
three-ground-plane design and perforated moats with 3.6\,$\mu$m
spacing in the two-ground-plane design performed equally well. Small
de facto moats around the thru-vias in the three-ground-plane design,
with maximum dimension of 5\,$\mu$m, did not trap flux with long moats
nearby. Even with the best moat design the circuits required a slow
cool-down rate. Rapid cool-down reportedly produces thermal gradients
leading to EMF currents and magnetic fields in the package
\cite{bermon1983moat}. Since the vortices that form in the ground
plane are only large and mobile in a narrow window above the freezing
temperature $T_f \approx T_c-15$\,mK, the required cool-down rate may
also indicate the time scale for the last vortex to move into the moat.


In conclusion, superconducting digital logic is scalable to increased
levels of integration with the development of Reciprocal Quantum Logic
and improved fabrication capabilities at sub-micron representing an
advance of five process nodes over previously established
technology. We report that flux trapping does not appear to be an
insurmountable limitation for large-scale superconductor integrated
circuits at advanced process nodes. Based on this success, we conclude
that further efforts are warranted to scale the technology yet
further.

\begin{acknowledgments}

  The authors acknowledge communication with Paul Bunyk, Vasili
  Semenov, and Sergey Tolpygo regarding effective moat geometry, Yuri
  Polyakov regarding the test fixture, and Mark Gouker regarding the
  strategy for circuit test.

\end{acknowledgments}

\nocite{*}
\bibliography{flux_rev}

\begin{thebibliography}{24}
\expandafter\ifx\csname natexlab\endcsname\relax\def\natexlab#1{#1}\fi
\expandafter\ifx\csname bibnamefont\endcsname\relax
  \def\bibnamefont#1{#1}\fi
\expandafter\ifx\csname bibfnamefont\endcsname\relax
  \def\bibfnamefont#1{#1}\fi
\expandafter\ifx\csname citenamefont\endcsname\relax
  \def\citenamefont#1{#1}\fi
\expandafter\ifx\csname url\endcsname\relax
  \def\url#1{\texttt{#1}}\fi
\expandafter\ifx\csname urlprefix\endcsname\relax\def\urlprefix{URL }\fi
\providecommand{\bibinfo}[2]{#2}
\providecommand{\eprint}[2][]{\url{#2}}

\bibitem[{\citenamefont{Abelson and Kerber}(2004)}]{abelson2004superconductor}
\bibinfo{author}{\bibfnamefont{L.~A.} \bibnamefont{Abelson}} \bibnamefont{and}
  \bibinfo{author}{\bibfnamefont{G.~L.} \bibnamefont{Kerber}},
  \bibinfo{journal}{Proceedings of the IEEE} \textbf{\bibinfo{volume}{92}},
  \bibinfo{pages}{1517} (\bibinfo{year}{2004}).

\bibitem[{\citenamefont{Herr et~al.}(2011)\citenamefont{Herr, Herr, Oberg, and
  Ioannidis}}]{herr2011ultra}
\bibinfo{author}{\bibfnamefont{Q.~P.} \bibnamefont{Herr}},
  \bibinfo{author}{\bibfnamefont{A.~Y.} \bibnamefont{Herr}},
  \bibinfo{author}{\bibfnamefont{O.~T.} \bibnamefont{Oberg}}, \bibnamefont{and}
  \bibinfo{author}{\bibfnamefont{A.~G.} \bibnamefont{Ioannidis}},
  \bibinfo{journal}{Journal of Applied Physics} \textbf{\bibinfo{volume}{109}},
  \bibinfo{pages}{103903} (\bibinfo{year}{2011}).

\bibitem[{\citenamefont{Herr et~al.}(2013)\citenamefont{Herr, Herr, Oberg,
  Naaman, Przybysz, Borodulin, and Shauck}}]{herr20138}
\bibinfo{author}{\bibfnamefont{A.~Y.} \bibnamefont{Herr}},
  \bibinfo{author}{\bibfnamefont{Q.~P.} \bibnamefont{Herr}},
  \bibinfo{author}{\bibfnamefont{O.~T.} \bibnamefont{Oberg}},
  \bibinfo{author}{\bibfnamefont{O.}~\bibnamefont{Naaman}},
  \bibinfo{author}{\bibfnamefont{J.~X.} \bibnamefont{Przybysz}},
  \bibinfo{author}{\bibfnamefont{P.}~\bibnamefont{Borodulin}},
  \bibnamefont{and} \bibinfo{author}{\bibfnamefont{S.~B.}
  \bibnamefont{Shauck}}, \bibinfo{journal}{Journal of Applied Physics}
  \textbf{\bibinfo{volume}{113}}, \bibinfo{pages}{033911}
  (\bibinfo{year}{2013}).

\bibitem[{\citenamefont{Johnson et~al.}(2010)\citenamefont{Johnson, Bunyk,
  Maibaum, Tolkacheva, Berkley, Chapple, Harris, Johansson, Lanting, and
  Perminov}}]{johnson2010scalable}
\bibinfo{author}{\bibfnamefont{M.}~\bibnamefont{Johnson}},
  \bibinfo{author}{\bibfnamefont{P.}~\bibnamefont{Bunyk}},
  \bibinfo{author}{\bibfnamefont{F.}~\bibnamefont{Maibaum}},
  \bibinfo{author}{\bibfnamefont{E.}~\bibnamefont{Tolkacheva}},
  \bibinfo{author}{\bibfnamefont{A.}~\bibnamefont{Berkley}},
  \bibinfo{author}{\bibfnamefont{E.}~\bibnamefont{Chapple}},
  \bibinfo{author}{\bibfnamefont{R.}~\bibnamefont{Harris}},
  \bibinfo{author}{\bibfnamefont{J.}~\bibnamefont{Johansson}},
  \bibinfo{author}{\bibfnamefont{T.}~\bibnamefont{Lanting}}, \bibnamefont{and}
  \bibinfo{author}{\bibfnamefont{I.}~\bibnamefont{Perminov}},
  \bibinfo{journal}{Superconductor Science and Technology}
  \textbf{\bibinfo{volume}{23}}, \bibinfo{pages}{065004}
  (\bibinfo{year}{2010}).

\bibitem[{\citenamefont{Tolpygo et~al.}(2015)\citenamefont{Tolpygo, Bolkhovsky,
  Weir, Johnson, Gouker, and Oliver}}]{tolpygo2015fabrication}
\bibinfo{author}{\bibfnamefont{S.~K.} \bibnamefont{Tolpygo}},
  \bibinfo{author}{\bibfnamefont{V.}~\bibnamefont{Bolkhovsky}},
  \bibinfo{author}{\bibfnamefont{T.~J.} \bibnamefont{Weir}},
  \bibinfo{author}{\bibfnamefont{L.~M.} \bibnamefont{Johnson}},
  \bibinfo{author}{\bibfnamefont{M.~A.} \bibnamefont{Gouker}},
  \bibnamefont{and} \bibinfo{author}{\bibfnamefont{W.~D.}
  \bibnamefont{Oliver}}, \bibinfo{journal}{Applied Superconductivity, IEEE
  Transactions on} \textbf{\bibinfo{volume}{25}}, \bibinfo{pages}{1101312}
  (\bibinfo{year}{2015}).

\bibitem[{\citenamefont{Bermon and Gheewala}(1983)}]{bermon1983moat}
\bibinfo{author}{\bibfnamefont{S.}~\bibnamefont{Bermon}} \bibnamefont{and}
  \bibinfo{author}{\bibfnamefont{T.}~\bibnamefont{Gheewala}},
  \bibinfo{journal}{Magnetics, IEEE Transactions on}
  \textbf{\bibinfo{volume}{19}}, \bibinfo{pages}{1160} (\bibinfo{year}{1983}).

\bibitem[{\citenamefont{Nagasawa et~al.}(1995)\citenamefont{Nagasawa, Numata,
  Kato, and Tahara}}]{nagasawa1995evaluation}
\bibinfo{author}{\bibfnamefont{S.}~\bibnamefont{Nagasawa}},
  \bibinfo{author}{\bibfnamefont{H.}~\bibnamefont{Numata}},
  \bibinfo{author}{\bibfnamefont{C.}~\bibnamefont{Kato}}, \bibnamefont{and}
  \bibinfo{author}{\bibfnamefont{S.}~\bibnamefont{Tahara}},
  \bibinfo{journal}{Extended Abstracts of 5th International Superconductive
  Electronics Conference} \textbf{\bibinfo{volume}{5}}, \bibinfo{pages}{192}
  (\bibinfo{year}{1995}).

\bibitem[{\citenamefont{Fujiwara et~al.}(2009)\citenamefont{Fujiwara, Nagasawa,
  Hashimoto, Hidaka, Yoshikawa, Tanaka, Akaike, Fujimaki, Takagi, and
  Takagi}}]{fujiwara2009research}
\bibinfo{author}{\bibfnamefont{K.}~\bibnamefont{Fujiwara}},
  \bibinfo{author}{\bibfnamefont{S.}~\bibnamefont{Nagasawa}},
  \bibinfo{author}{\bibfnamefont{Y.}~\bibnamefont{Hashimoto}},
  \bibinfo{author}{\bibfnamefont{M.}~\bibnamefont{Hidaka}},
  \bibinfo{author}{\bibfnamefont{N.}~\bibnamefont{Yoshikawa}},
  \bibinfo{author}{\bibfnamefont{M.}~\bibnamefont{Tanaka}},
  \bibinfo{author}{\bibfnamefont{H.}~\bibnamefont{Akaike}},
  \bibinfo{author}{\bibfnamefont{A.}~\bibnamefont{Fujimaki}},
  \bibinfo{author}{\bibfnamefont{K.}~\bibnamefont{Takagi}}, \bibnamefont{and}
  \bibinfo{author}{\bibfnamefont{N.}~\bibnamefont{Takagi}},
  \bibinfo{journal}{Applied Superconductivity, IEEE Transactions on}
  \textbf{\bibinfo{volume}{19}}, \bibinfo{pages}{603} (\bibinfo{year}{2009}).

\bibitem[{\citenamefont{Jeffery et~al.}(1995)\citenamefont{Jeffery, Van~Duzer,
  Kirtley, and Ketchen}}]{jeffery1995magnetic}
\bibinfo{author}{\bibfnamefont{M.}~\bibnamefont{Jeffery}},
  \bibinfo{author}{\bibfnamefont{T.}~\bibnamefont{Van~Duzer}},
  \bibinfo{author}{\bibfnamefont{J.}~\bibnamefont{Kirtley}}, \bibnamefont{and}
  \bibinfo{author}{\bibfnamefont{M.}~\bibnamefont{Ketchen}},
  \bibinfo{journal}{Applied Physics Letters} \textbf{\bibinfo{volume}{67}},
  \bibinfo{pages}{1769} (\bibinfo{year}{1995}).

\bibitem[{\citenamefont{Robertazzi et~al.}(1997)\citenamefont{Robertazzi,
  Siddiqi, and Mukhanov}}]{robertazzi1997flux}
\bibinfo{author}{\bibfnamefont{R.}~\bibnamefont{Robertazzi}},
  \bibinfo{author}{\bibfnamefont{I.}~\bibnamefont{Siddiqi}}, \bibnamefont{and}
  \bibinfo{author}{\bibfnamefont{O.}~\bibnamefont{Mukhanov}},
  \bibinfo{journal}{Applied Superconductivity, IEEE Transactions on}
  \textbf{\bibinfo{volume}{7}}, \bibinfo{pages}{3164} (\bibinfo{year}{1997}).

\bibitem[{\citenamefont{Polyakov et~al.}(2011)\citenamefont{Polyakov, Semenov,
  and Tolpygo}}]{polyakov20113d}
\bibinfo{author}{\bibfnamefont{Y.~A.} \bibnamefont{Polyakov}},
  \bibinfo{author}{\bibfnamefont{V.~K.} \bibnamefont{Semenov}},
  \bibnamefont{and} \bibinfo{author}{\bibfnamefont{S.~K.}
  \bibnamefont{Tolpygo}}, \bibinfo{journal}{Applied Superconductivity, IEEE
  Transactions on} \textbf{\bibinfo{volume}{21}}, \bibinfo{pages}{724}
  (\bibinfo{year}{2011}).

\bibitem[{\citenamefont{Terai et~al.}(2007)\citenamefont{Terai, Tanaka,
  Yamanashi, Hashimoto, Fujimaki, Yoshikawa, and Wang}}]{terai2007diagnostic}
\bibinfo{author}{\bibfnamefont{H.}~\bibnamefont{Terai}},
  \bibinfo{author}{\bibfnamefont{M.}~\bibnamefont{Tanaka}},
  \bibinfo{author}{\bibfnamefont{Y.}~\bibnamefont{Yamanashi}},
  \bibinfo{author}{\bibfnamefont{Y.}~\bibnamefont{Hashimoto}},
  \bibinfo{author}{\bibfnamefont{A.}~\bibnamefont{Fujimaki}},
  \bibinfo{author}{\bibfnamefont{N.}~\bibnamefont{Yoshikawa}},
  \bibnamefont{and} \bibinfo{author}{\bibfnamefont{Z.}~\bibnamefont{Wang}},
  \bibinfo{journal}{Applied Superconductivity, IEEE Transactions on}
  \textbf{\bibinfo{volume}{17}}, \bibinfo{pages}{422} (\bibinfo{year}{2007}).

\bibitem[{\citenamefont{Semenov et~al.}(1999)\citenamefont{Semenov, Polyakov,
  and Chao}}]{semenov1999extraction}
\bibinfo{author}{\bibfnamefont{V.}~\bibnamefont{Semenov}},
  \bibinfo{author}{\bibfnamefont{Y.~A.} \bibnamefont{Polyakov}},
  \bibnamefont{and} \bibinfo{author}{\bibfnamefont{W.}~\bibnamefont{Chao}},
  \bibinfo{journal}{Applied Superconductivity, IEEE Transactions on}
  \textbf{\bibinfo{volume}{9}}, \bibinfo{pages}{4030} (\bibinfo{year}{1999}).

\bibitem[{\citenamefont{Polyakov et~al.}(2007)\citenamefont{Polyakov, Narayana,
  and Semenov}}]{polyakov2007flux}
\bibinfo{author}{\bibfnamefont{Y.}~\bibnamefont{Polyakov}},
  \bibinfo{author}{\bibfnamefont{S.}~\bibnamefont{Narayana}}, \bibnamefont{and}
  \bibinfo{author}{\bibfnamefont{V.~K.} \bibnamefont{Semenov}},
  \bibinfo{journal}{Applied Superconductivity, IEEE Transactions on}
  \textbf{\bibinfo{volume}{17}}, \bibinfo{pages}{520} (\bibinfo{year}{2007}).

\bibitem[{\citenamefont{Narayana et~al.}(2009)\citenamefont{Narayana, Polyakov,
  and Semenov}}]{narayana2009evaluation}
\bibinfo{author}{\bibfnamefont{S.}~\bibnamefont{Narayana}},
  \bibinfo{author}{\bibfnamefont{Y.}~\bibnamefont{Polyakov}}, \bibnamefont{and}
  \bibinfo{author}{\bibfnamefont{V.~K.} \bibnamefont{Semenov}},
  \bibinfo{journal}{Applied Superconductivity, IEEE Transactions on}
  \textbf{\bibinfo{volume}{19}}, \bibinfo{pages}{640} (\bibinfo{year}{2009}).

\bibitem[{\citenamefont{Semenov et~al.}(2015)\citenamefont{Semenov, Polyakov,
  and Tolpygo}}]{semenov2015new}
\bibinfo{author}{\bibfnamefont{V.~K.} \bibnamefont{Semenov}},
  \bibinfo{author}{\bibfnamefont{Y.}~\bibnamefont{Polyakov}}, \bibnamefont{and}
  \bibinfo{author}{\bibfnamefont{S.~K.} \bibnamefont{Tolpygo}},
  \bibinfo{journal}{Applied Superconductivity, IEEE Transactions on}
  \textbf{\bibinfo{volume}{25}}, \bibinfo{pages}{1301507}
  (\bibinfo{year}{2015}).

\bibitem[{\citenamefont{Stan et~al.}(2004)\citenamefont{Stan, Field, and
  Martinis}}]{stan2004critical}
\bibinfo{author}{\bibfnamefont{G.}~\bibnamefont{Stan}},
  \bibinfo{author}{\bibfnamefont{S.~B.} \bibnamefont{Field}}, \bibnamefont{and}
  \bibinfo{author}{\bibfnamefont{J.~M.} \bibnamefont{Martinis}},
  \bibinfo{journal}{Physical Review Letters} \textbf{\bibinfo{volume}{92}},
  \bibinfo{pages}{097003} (\bibinfo{year}{2004}).

\bibitem[{\citenamefont{Kuit et~al.}(2008)\citenamefont{Kuit, Kirtley, Van
  Der~Veur, Molenaar, Roesthuis, Troeman, Clem, Hilgenkamp, Rogalla, and
  Flokstra}}]{kuit2008vortex}
\bibinfo{author}{\bibfnamefont{K.}~\bibnamefont{Kuit}},
  \bibinfo{author}{\bibfnamefont{J.}~\bibnamefont{Kirtley}},
  \bibinfo{author}{\bibfnamefont{W.}~\bibnamefont{Van Der~Veur}},
  \bibinfo{author}{\bibfnamefont{C.}~\bibnamefont{Molenaar}},
  \bibinfo{author}{\bibfnamefont{F.}~\bibnamefont{Roesthuis}},
  \bibinfo{author}{\bibfnamefont{A.}~\bibnamefont{Troeman}},
  \bibinfo{author}{\bibfnamefont{J.}~\bibnamefont{Clem}},
  \bibinfo{author}{\bibfnamefont{H.}~\bibnamefont{Hilgenkamp}},
  \bibinfo{author}{\bibfnamefont{H.}~\bibnamefont{Rogalla}}, \bibnamefont{and}
  \bibinfo{author}{\bibfnamefont{J.}~\bibnamefont{Flokstra}},
  \bibinfo{journal}{Physical Review B} \textbf{\bibinfo{volume}{77}},
  \bibinfo{pages}{134504} (\bibinfo{year}{2008}).

\bibitem[{\citenamefont{Pearl}(1964)}]{pearl1964current}
\bibinfo{author}{\bibfnamefont{J.}~\bibnamefont{Pearl}},
  \bibinfo{journal}{Applied Physics Letters} \textbf{\bibinfo{volume}{5}},
  \bibinfo{pages}{65} (\bibinfo{year}{1964}).

\bibitem[{\citenamefont{Clem}(1998)}]{clem1998vortex}
\bibinfo{author}{\bibfnamefont{J.~R.} \bibnamefont{Clem}},
  \bibinfo{journal}{APS March Meeting Abstracts} \textbf{\bibinfo{volume}{1}},
  \bibinfo{pages}{3606} (\bibinfo{year}{1998}).

\bibitem[{\citenamefont{Likharev}(1972)}]{likharev1972formation}
\bibinfo{author}{\bibfnamefont{K.}~\bibnamefont{Likharev}},
  \bibinfo{journal}{Sov. Radiophys} \textbf{\bibinfo{volume}{14}},
  \bibinfo{pages}{722} (\bibinfo{year}{1972}).

\bibitem[{\citenamefont{Maksimova}(1998)}]{maksimova1998mixed}
\bibinfo{author}{\bibfnamefont{G.}~\bibnamefont{Maksimova}},
  \bibinfo{journal}{Physics of the Solid State} \textbf{\bibinfo{volume}{40}},
  \bibinfo{pages}{1607} (\bibinfo{year}{1998}).

\bibitem[{\citenamefont{Tolpygo et~al.}(2014)\citenamefont{Tolpygo, Bolkhovsky,
  Johnson, Oliver, and Gouker}}]{tolpygo2014roadmap}
\bibinfo{author}{\bibfnamefont{S.~K.} \bibnamefont{Tolpygo}},
  \bibinfo{author}{\bibfnamefont{V.}~\bibnamefont{Bolkhovsky}},
  \bibinfo{author}{\bibfnamefont{L.~M.} \bibnamefont{Johnson}},
  \bibinfo{author}{\bibfnamefont{W.~D.} \bibnamefont{Oliver}},
  \bibnamefont{and} \bibinfo{author}{\bibfnamefont{M.~A.}
  \bibnamefont{Gouker}}, \bibinfo{journal}{Superconductivity News Forum Global
  Edition} \textbf{\bibinfo{volume}{30}}, \bibinfo{pages}{STP402}
  (\bibinfo{year}{2014}).

\bibitem[{\citenamefont{Likharev and Semenov}(1991)}]{likharev1991rsfq}
\bibinfo{author}{\bibfnamefont{K.~K.} \bibnamefont{Likharev}} \bibnamefont{and}
  \bibinfo{author}{\bibfnamefont{V.~K.} \bibnamefont{Semenov}},
  \bibinfo{journal}{Applied Superconductivity, IEEE Transactions on}
  \textbf{\bibinfo{volume}{1}}, \bibinfo{pages}{3} (\bibinfo{year}{1991}).

\end{thebibliography}

\end{document}